\documentclass[useAMS,usenatbib]{mn2e}
\usepackage{graphicx,epstopdf}

\title{Radial mixing and the transition between the thick and thin Galactic discs}
\author[M. Haywood]
       {M. Haywood\thanks{email : Misha.Haywood@obspm.fr} \\
        GEPI, Observatoire de  Paris, F-92195 Meudon Cedex,France}
\date{Accepted.
      Received ;
      in original form }
\pagerange{\pageref{firstpage}--\pageref{lastpage}}
\pubyear{2006}          
\begin{document}         

\maketitle
\begin{abstract} The analysis of the kinematics of solar neighbourhood
stars shows that  the low and high metallicity tails  of the thin disc
are populated by objects which orbital properties suggest an origin in
the outer and inner galactic disc, respectively.  Signatures of radial
migration are identified  in various recent samples, and  are shown to
be   responsible  for   the   high  metallicity   dispersion  in   the
age-metallicity distribution.  Most importantly,  it is shown that the
population   of   low  metallicity   wanderers   of   the  thin   disc
(-0.7$<$[Fe/H]$<$-0.3 dex) is also responsible for the apparent hiatus
in  metallicity with  the thick  disc (which  terminal  metallicity is
about -0.2 dex). It implies that the thin disc at the solar circle has
started to  form stars at about  this same metallicity.   This is also
consistent  with  the  fact  that  'transition'  objects,  which  have
$\alpha$-element abundance intermediate between  that of the thick and
thin  discs,  are  found  in  the range  [-0.4,-0.2]  dex.   Once  the
metal-poor  thin  disc  stars  are  recognised for  what  they  are  -
wanderers from  the outer thin disc  - the parenthood  between the two
discs can be identified on  stars genuinely formed at the solar circle
through  an  evolutionary  sequence  in  [$\alpha$/Fe]  and  [Fe/H]  .
Another  consequence is  that stars  that can  be considered  as truly
resulting  of  the chemical  evolution  at  the  solar circle  have  a
metallicity  restricted to  about [-0.2,+0.2]  dex, confirming  an old
idea that most  chemical evolution in the Milky  Way have preceded the
thin disc formation.
\end{abstract}

 
\section{Introduction} The  Galactic thick  disc is, according  to the
local record, a population  with characteristics neighbouring those of
the thin disc  : it is rotationally supported,  with a higher internal
kinematic   dispersion,  slightly  metal-poor   with  an   overlap  in
metallicity with  this population. It is however  possibly much older,
and  shows clear  discontinuities with  the thin  disc, as  visible in
particular on chemical data.  Several  scenarios of the origin of this
population  have been  put forward  to explain  these characteristics,
among which the  following three are often evoked:  (1) the thick disc
is an accreted population, (2) it is the first phase of the thin disc,
heated up by  an interaction episode with another  galaxy (3) finally,
it  could  be  the result  of  multiple,  early  mergers of  gas  rich
subsystems from which stars of the thick disc would have formed.  This
last suggestion  has been made by  Brook et al.  (2007) from numerical
simulations where they apparently succeed to produce a population with
the main properties similar to those  of the thick disc.  In the first
scenario, the properties  of the thick disc are  those of the accreted
population.  This  is  a   difficulty,  because  while  the  range  of
properties found in Milky Way satellites is rather large, none of them
approach those of the thick disc.  It is doubtful in particular that a
small  entity could have  produced, at  an early  epoch, stars  with a
relatively   high   level   of   $\alpha$-elements  as   observed   on
solar-neighbourhood thick disc objects.  In the second scenario, it is
not entirely clear what exactly  occurs in the transition phase (which
may  have  lasted  several  Gyr)  between  the  interaction  with  the
satellite and the beginning of  the thin disc phase, and in particular
to what discontinuities it may  give rise.  While it is uncertain what
may differentiate the resulting thick  disc of scenario (2) and (3), a
kind of parenthood is certainly  expected with the thin disc, but this
is unlikely in scenario (1).

In  the recent  years, detailed  spectroscopic data  have demonstrated
that  the  two  discs   are  apparently  distinct  in  their  chemical
properties, with two main  differences. First, at a given metallicity,
the $\alpha$-element content of the thick disc is offset by about +0.1
dex from  the thin disc.  Second, there is  a hiatus of about  0.5 dex
between the metal-rich stars of the thick disc (at [Fe/H]$\approx$-0.2
dex)  and metal-poor stars  of the  thin disc  (at [Fe/H]$\approx$-0.7
dex).   It  has been  suggested  that  these  two sequences  could  be
evolutionary connected, (Bensby  et al. 2003; Reddy et  al. 2003), the
metallicity hiatus  between the two populations  being the consequence
of a  phase of low  star formation, perhaps  due to the  exhaustion of
gas,  followed by  a  new  infall episode,  which  would decrease  the
metallicity from which the new, thin disc stars, form.

However,  this   peculiar  interpretation  contradicts   the  observed
behaviour  of  the age-metallicity  distribution  (Haywood, 2006,  and
Fig. 1,  next section) which shows  that, on average,  the oldest thin
disc  stars  have  a  metallicity  of  about  -0.2  dex  --  not  -0.7
dex. Meanwhile,  thin disc stars this metal-poor  are not specifically
old, being found  also at intermediate ages (starting  at 2-3 Gyr).  A
different interpretation,  suggested in Haywood (2006)  and studied in
more detail  in the present paper,  is that the  metallicity spread of
thin disc stars at [-0.3,-0.7]  dex is a consequence of radial mixing,
stars more  metal-poor than  about -0.3 dex  being intruders  from the
outer disc.  It has been proposed (Grenon, 1972) that super metal-rich
([Fe/H]$>$0.2 dex) stars found  in the solar neighbourhood are objects
that have migrated from the inner disc.  It is suggested here that the
same phenomenon explains the presence of the most metal-poor thin disc
stars at the  solar radius, these being wanderers  from the outer thin
disc.  If correct, it would  imply that the metallicity hiatus between
the  two populations  would not  be an  effect of  the  local chemical
evolution, but a  consequence of radial mixing in  the thin disc.  The
present study focuses on  these discontinuities, showing that they are
only apparent, and that a real evolutionary link between the two discs
does exist.  We  emphasize that the present results  are obtained from
solar neighbourhood stars.

The present  paper is organised as  follows. In the  section below, we
start  by looking at  the signatures  of radial  mixing in  samples of
stars of the  solar vicinity.  We then analyse  the transition between
the  thick and  thin disc  in the  light of  these results,  and focus
particularly  on  the  discontinuities  in [$\alpha$/Fe]  and  [Fe/H],
arguing they  are only  apparent.  We conclude  with a summary  of our
results and a discussion on the status of the thick disc.

\section{Observational signatures of Radial mixing}

Radial migration of  stars in the galactic disc  has been suggested to
be an important ingredient of  our understanding of the local galactic
populations  since  the seventies  (Grenon,  1972),  in particular  to
justify the now  compelling evidence of metal-rich stars  in the solar
neighbourhood.  A  variety of processes  have been invoked  to explain
radial  mixing,  including the  diffusion  of  stars  on their  orbits
because   of  various   irregularities  in   the   galactic  potential
(quantified by Wielen (1977) and  Wielen et al.  (1996)), or caused by
the passage of  either a transient single spiral  pattern (Sellwood \&
Binnney, 2002),  or multiple long  lived or transient  spiral patterns
(Minchev \& Quillen (2006), De Simone et al. (2004)).

There   are  two  expected   signatures  of   radial  mixing   on  the
metallicities of  thin disc stars in the  solar neighbourhood.  Radial
migration being a secular process, its effects are expected to be more
or less correlated with age: the oldest stars are allowed to come from
more  distant regions,  hence from  regions where  metallicity  can be
increasingly  different  from  the   mean  metallicity  of  the  solar
neighbourhood. Therefore, if radial mixing is the prime contributor to
the  dispersion  of  metallicities  at  the solar  circle,  we  expect
dispersion to increase with age.  However, other mechanisms can induce
a similar effect. For example, infalling gas from the halo could alter
the homogeneity of  the ISM, while the amplitude  of this effect could
be  a  diminishing  function  of  time,  correlated  with  a  possibly
decreasing amount  of gas in  the halo 'reservoir'.  The  second, more
specific, signature of radial mixing, is due to the radial metallicity
gradient: because  stars born  in the inner  disc are  more metal-rich
than stars  born in the  outer disc, we  may search for  signatures of
these opposite  provenance in the orbital behaviour  of the metal-rich
and metal-poor solar neighbourhood stars.

\subsection{The increasing dispersion of metallicity with age}

\begin{figure*}
\includegraphics[width=18cm]{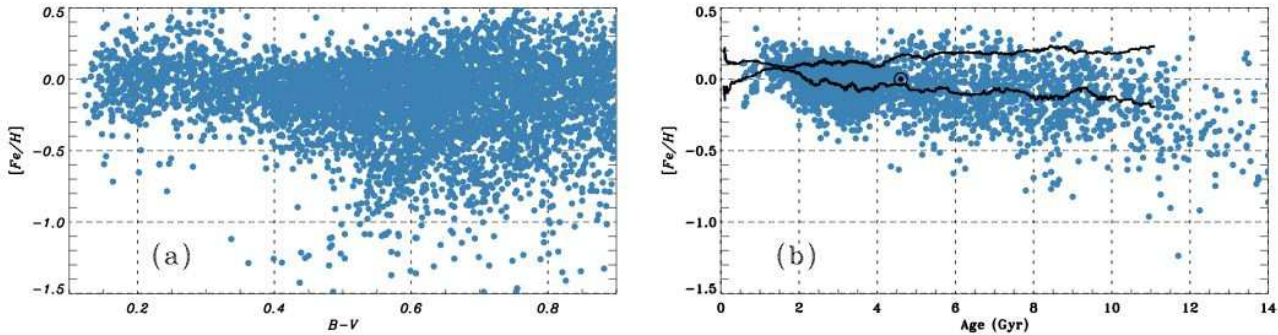}
\caption{Illustration of the increasing metallicity dispersion with age in two different samples: 
(a) The (B-V-[Fe/H]) distribution. Metallicity has been derived from Geneva photometry and (b) age-metallicity distribution.
The curves are the mean and dispersion of metallicity as a function of age.
Metallicities are from Str\"omgren photometry and ages have been derived as mentioned in the text.
The position of the Sun if also indicated.
The two panels illustrate the same tendency of an increase in the metallicity dispersion with age, which
can be interpreted as the pollution of the solar neighbourhood by stars coming from both the outer and inner disc. 
}
\label{rmeffect}
\end{figure*}

We  plot  on Fig.  \ref{rmeffect}ab  two  different  samples of  solar
neighbourhood stars.  The first one  is made of 5600 stars within 75pc
and shows metallicity derived  through Geneva photometry as a function
of B-V. The  second sample contains about 3300  stars with metallicity
from Str\"{o}mgren  photometry. This sample is used  in Haywood (2006)
to  derived the age-metallicity  distribution.  However,  a noticeable
difference with  Haywood (2006) is that  we derived new  ages with the
method developed  by J{\o}rgensen \& Lindegren (2005).   The method is
based on the  evaluation of an age probability  density function (pdf)
for each star,  the adopted age being determined from  the mode of the
pdf.  All the  details are given in J{\o}rgensen  \& Lindegren (2005).
We used  the set  of isochrones  from Demarque et  al. (2004)  and the
atmospheric  parameters are  those of  Haywood (2006).   The resulting
age-[Fe/H] distribution  has the same main characteristics  as the one
found in Haywood (2006).

The expected  trend of increasing  dispersion with age is  observed on
both plots  (implicitly in panel (a)).   The upper curve  on panel (b)
quantifies  the increase  in metallicity  dispersion: starting  at 0.1
dex,  the  dispersion increases  to  0.25  dex  for older  ages.   The
widening  of  the  metallicity  distribution with  increasing  B-V  is
apparent in  panel (a).   The proportion of  old disc  stars increases
from B-V=0.3  to 0.6, bringing more objects  with wider metallicities,
and, possibly, of more distant  origin.  Note that this is independent
of  age determination  and  is a  robust  result :  the dispersion  in
metallicity is not  uniform at all ages but  increases from younger to
older generation  of stars.  Young  stars (0.35$<$B-V$<$0.45) indicate
that  the  most metal-rich  objects  presently  forming  in the  solar
neighbourhood have  a metallicity of  about [Fe/H]$\approx$0.1-0.2 dex
(the  clump at  B-V$<$0.3 is  made of  metallic and  peculiar A  and F
stars, and is not considered here), while the inclusion of redder, and
presumably older  stars in the sample brings  more metal-rich objects.
If it is admitted that the metallicity of young stars is the end-point
of the  local chemical  evolution, it is  natural to suppose  that the
metal-rich,  older stars seen  at B-V$>$0.5  and [Fe/H]$>$0.2  dex are
intruders of different galactic origin.

The role of radial mixing  to explain the existence of metal-poor thin
disc stars is  usually not invoked, because these  are usually thought
to   stem   from  the   local   chemical   evolution.   However,   the
age-metallicity relation  in panel  (b) suggest another  picture.  The
average metallicity of old thin disc stars is [Fe/H]$\approx$-0.2 dex,
and there is  no hint of a transition between  two populations at ages
8-12  Gyr   and  [Fe/H]=-0.7  dex.    On  the  contrary,   stars  with
metallicities -0.7$<$[Fe/H]$<$-0.3  dex are spread at  various ages in
the  interval [2-3,  8-10]  Gyr,  which imply  they  are not  suitable
candidates  for  being  transition  objects  between  the  two  discs.
Another noticeable characteristic  from Fig. \ref{rmeffect}(b) is that
none of these objects have age less than 2 Gyr, which means that stars
at  such metallicities  do not  originate from  - or  are  not forming
anymore at - the solar neighbourhood.

Finally, the metallicity of the Sun  relative to stars of the same age
in  the solar  neighbourhood (differences  of 0.15-0.2  dex  have been
claimed)  has  been invoked  as  evidence  of  radial mixing  (see  in
particular  Wielen  et  al.  (1996)).   The location  of  the  Sun  on
Fig. \ref{rmeffect}(b) is  indicated by its symbol, and  shows that it
is offset  by less  than 0.1  dex ($\approx$ 0.05  dex) from  the mean
metallicity  of  stars  at  the  same age,  seriously  weakening  this
argument.  In  the section  below, we search  for other  signatures on
local samples of stars.

\subsection{Systematic variations of orbital parameters with metallicity}

If we assume that stars in  the metal-poor and metal-rich tails of the
solar neighbourhood  metallicity distribution are  mostly contaminants
passing through the solar circle as  a result of radial mixing, we may
expect to detect signatures of this migration process in the kinematic
and orbital data of solar neighbourhood stars.

\subsubsection{A photometric sample : Haywood (2006)}

Our first sample has been used  in the previous section and in Haywood
(2006)  to  determine  the  age-metallicity  relation.   No  kinematic
criterion has been  used to select the stars.  We cross-identified our
sample with the  CGS catalogue of Nordstr\"om et  al. (2004) to obtain
the kinematic  and orbital parameters  U,V,W and R$_{p}$,  R$_{a}$ for
about 3300 objects.

We  assigned  population  membership  probabilities according  to  the
method of Mishenina et al. (2004)  for all the stars in our sample. We
adopted the kinematic parameters for  the thin and thick discs and the
Hercules stream given by these authors.  The probabilities are derived
assuming that  the kinematic distributions of  the stellar populations
are Gaussians.   Relative densities have  been assumed to be  82\% for
the  thin disc,  10\% for  the  thick disc  and 8\%  for the  Hercules
stream, in agreement with current  estimates (see Famaey et al. (2005)
for example).  Each star has a calculated probability to belong to the
thin  disc (P$_{TND}$)  the  thick disc  (P$_{TKD}$)  or the  Hercules
stream  (P$_H$)  (the halo  has  been  neglected:  3 stars  only  have
[Fe/H]$<$-1.5).   The  detailed procedure  is  given  in Mishenina  et
al. (2004) (see  also Soubiran \& Girard, 2005),  and is not described
further.

\begin{figure}
\includegraphics{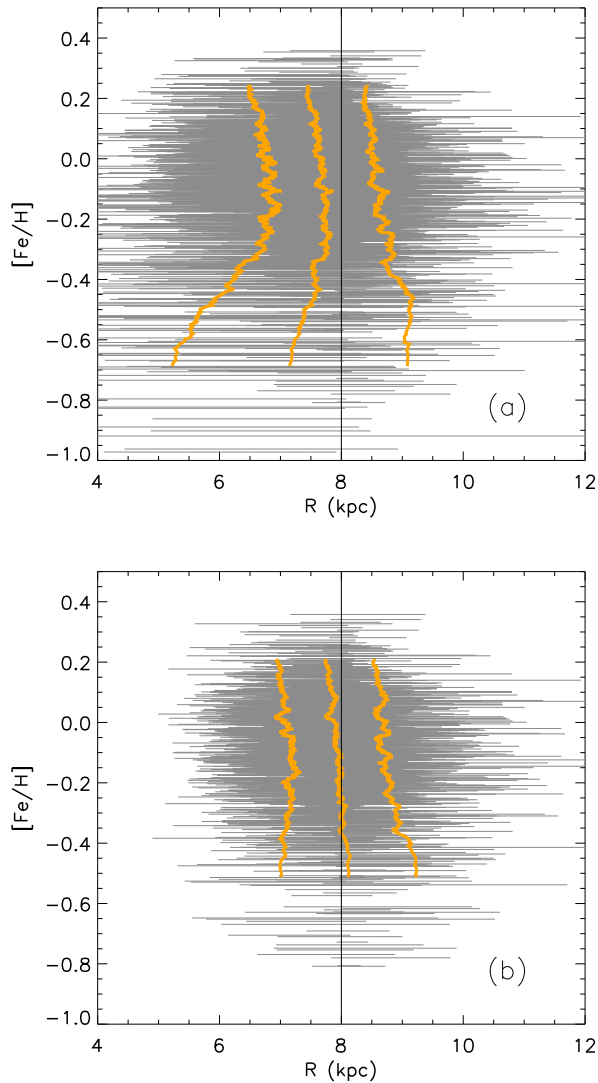}
\caption{(a) Radial amplitude (distance from R$_{min}$ to R$_{max}$) of the orbits of all 
the stars in the photometric sample as a function of metallicity. 
(b) Thin disc stars only, selected assuming a probability higher than 0.7.
The curves are the mean of R$_{p}$, R$_m$ and R$_{a}$, calculated on subsamples of 
100 (non-independent) objects. 
}
\label{rmthindisk}
\end{figure}

Figure \ref{rmthindisk}(a)  shows the radial  galactocentric amplitude
(R$_{p}$  to R$_{a}$)  of the  orbit  of the  stars as  a function  of
metallicity  for the entire  sample (a)  and the  thin disc  (b), this
population being selected with kinematic membership probability higher
than  0.7.  On  each plot,  curves represent  the mean  orbital radius
R$_m$  (defined as  (R$_{p}$+R$_{a}$)/2), calculated  on (overlapping)
subsamples of 100 points.  The whole sample (panel a) illustrates that
there is  a significant increase of the  galactocentric radius towards
lower metallicity  stars (from R$_m$=7.5  kpc at [Fe/H]=+0.25  to 7.75
kpc at [Fe/H]=-0.3  dex).  Below this limit, the  contamination by the
thick  disc makes the  mean galactocentric  radius to  decrease again.
Panel  2c shows  the thin  disc  stars only,  selected with  kinematic
probability higher  than 70\%.  There  is a clear visible  increase of
R$_{m}$  at lower  metallicities in  the thin  disc, with  R$_{p}$ and
R$_{a}$  being  similarly shifted.   It  illustrates  that, for  stars
belonging to the  thin disc, the orbits are  shifted towards the outer
disc  at lower  metallicities and  towards  the inner  disc at  higher
metallicities, with a change in  the mean R$_{m}$ limited to about 0.5
kpc on the metallicity interval (+0.2, -0.5) dex.

\subsubsection{A spectroscopic sample : Soubiran \& Girard (2005)}

There  is  a  possibility  that   the  increase  in  R$_{m}$  seen  on
Fig.  \ref{rmthindisk}b at  [Fe/H]$<$-0.4  dex is  partly  due to  the
kinematic  selection   of  disc   stars,  because  the   selection  of
P$_{TND}>$0.7  eliminates stars  that have  a  kinematics intermediate
between the thin and thick discs, and the adopted level at which stars
of each population should  be selected is somewhat arbitrary.  Another
way to discriminate thin and thick disc stars is to use their distinct
chemistry.  In many spectroscopic  studies of the solar neighbourhood,
the thin and thick discs  are often better separated in their chemical
properties   than  in   their  kinematics.    Figure  \ref{rmthindisk}
illustrates that R$_{m}$ varies as a function of metallicity even when
no   selection  on  the   population  is   made  (at   [Fe/H]$>$  -0.3
dex).  However, we  conducted an  additional test  to verify  that the
effect  measured at  low metallicities  is  not a  consequence of  the
kinematic selection.  In  order to do so, we  repeat the same analysis
as  above  but  on a  sample  for  which  [Mg/Fe] abundance  ratio  is
available and use it to select  thin disc star.  The new sample is the
compilation by  Soubiran \&  Girard (2005), which  contains kinematic,
orbital and chemical data for a  set of 743 stars. An inconvenience is
that part of  the separation in abundance ratios  between the thin and
thick  discs   visible  on  individual  samples  is   blurred  by  the
combination of several data sets of different origins, but some of the
information is certainly exploitable.   In order to have kinematic and
orbital parameters from the same  source as the photometric sample, we
cross-identified the compilation of Soubiran \& Girard (2005) with the
catalogue of Nordstr\"om  et al. (2004), leaving about  600 stars with
measured [Mg/Fe].

\begin{figure}
\includegraphics{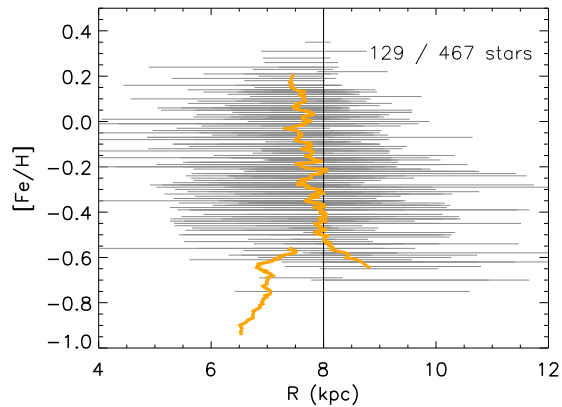}
\caption{The same as Figure \ref{rmthindisk}, with the sample from Soubiran \& Girard (2005)
as described in the text. 
The classification between thick (left curve) and thin (right curve) discs is made according to the criterion 
that [Mg/Fe]$>$0.2 dex and [Mg/Fe]$<$0.2 dex.
The curves are the mean of R$_m$ calculated on (non-independent) samples of 70 points.
}
\label{rminfeh2}
\end{figure}

Fig. \ref{rminfeh2} shows radial excursions of the stars as a function
of  metallicity. The separation  between the  two populations  is made
according to the  limit [Mg/Fe]=0.2 dex.  The curves  show the mean of
R$_{m}$  calculated  on subsamples  of  70  stars.   The behaviour  of
R$_{m}$ for the  thin disc population (right curve)  is similar to the
one   seen   on   the   photometric   sample,   with   the   kinematic
selection.  There is  a tendency  for an  increase of  R$_{m}$ towards
decreasing thin  disc metallicity,  confirming the possibility  of the
contamination  of the  vicinity by  wanderers of  the inner  and outer
disc.

\begin{figure}
\includegraphics{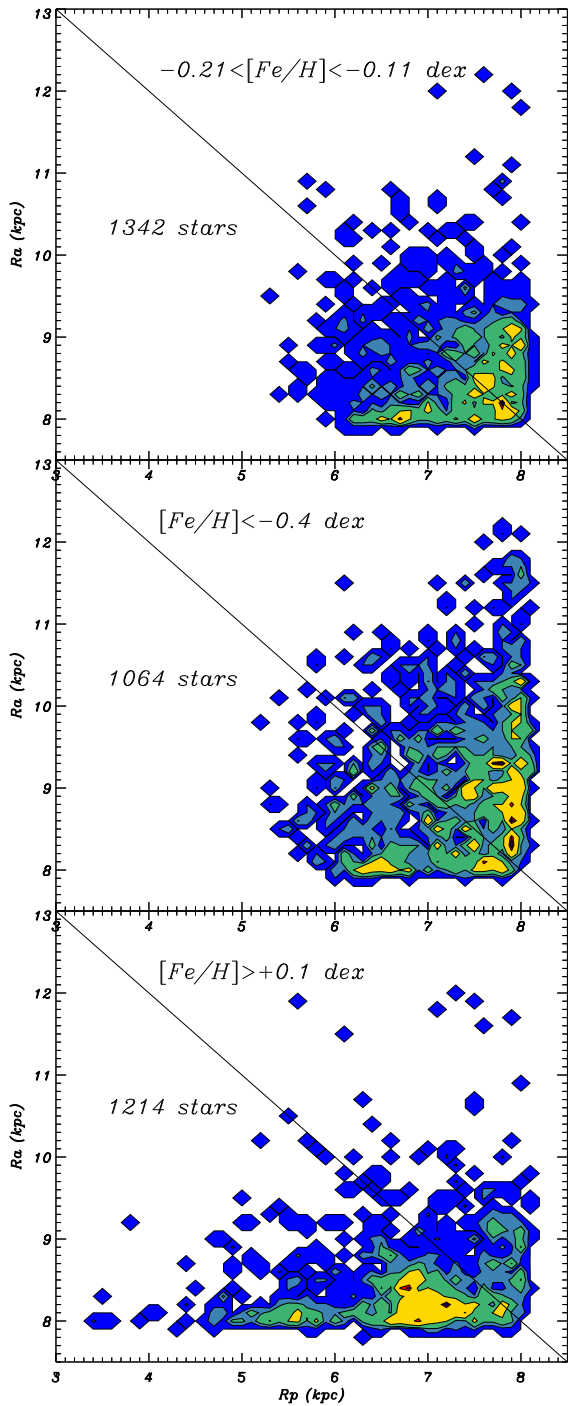}
\caption{The distribution of stars kinematically selected as thin disc 
stars in the GCS catalogue, on different metallicity intervals. 
The diagonal line is for a mean galactocentric radius of 8 kpc.
(a) Reference sample, for stars in a metallicity interval centred on the mode
of the metallicity distribution ($\pm$ 0.05 dex) of the GCS catalogue, selected assuming P$_{TND}>0.8$. 
(b) Thin disc stars (P$_{TND}>0.8$) with [Fe/H]$<$-0.4 dex.
(c) Stars with [Fe/H]$>$+0.1 dex. The Hyades stream is visible at
(6.7-7.0, 8.0-8.7)kpc. In this metallicity interval, we assumed there is no 
contamination by the thick disc (no selection on membership probability was made). 
}
\label{dfehnma}
\end{figure}

\subsubsection{The Geneva-Copenhagen Survey}

We now  consider a larger sample,  that of Nordstr\"om  et al. (2004),
which  provides  kinematic  and  orbital parameters  for  about  12000
objects.  The overall sample shows  the same trends as those evidenced
in the two previously studied data  sets, and we don't repeat the same
analysis.  However, these  trends can  be  seen even  more clearly  on
(R$_{p}$, R$_{a}$) distributions as  given by two subsamples separated
in  metallicities,  see  Fig.  \ref{dfehnma}.   Panel  (a)  gives  the
distribution  around   the  central  metallicity  of   the  sample  of
Nordstr\"om  et  al., which  is  about  -0.1 dex,  and  is  used as  a
reference   distribution.  The  difference   between  the   two  other
distributions,  plotted for [Fe/H]$<$-0.4  dex and  [Fe/H]$>$+0.1 dex,
can be seen on panel (b) and (c).  The metal-poor subsample is shifted
to  larger  apo  and  peri-centres, and  in  particular  significantly
extended  beyond 9.5 kpc.   The concentration  seen in  the metal-rich
subsample  at (R$_{p}$, R$_{a}$)=(6.6-7.2,8.1-8.6)  kpc is  the Hyades
stream.  Famaey et al. (2007) argued that the Hyades stream is a group
of stars  coincidentally brought together  from the inner disc  to the
solar  vicinity  by  the   effect  of  a  spiral  perturbation.   This
suggestion  fits well  within the  general picture  proposed  here and
contributes importantly to the detected effect.

The diagonal  line on each plot  indicates a mean orbital  radius of 8
kpc.  The symmetrical aspect of these plots is obvious: the metal-poor
thin  disc  populates preferentially  outer  orbits, while  metal-rich
stars  have a  net  tendency  to occupy  inner  orbits.  Although  the
differences  are clear,  there  is  a large  overlap  between the  two
distributions of panel (b) and  (c).  As an example, the Hyades stream
supposedly   has  no   stars  below   [Fe/H]$<$-0.5  dex   (Famaey  et
al. (2005)), but probably contaminates the sample at [Fe/H]$<$-0.4 dex
on Fig. \ref{dfehnma}(b).   In addition, while we do  expect to select
stars that  are preferentially on  outer orbits at  [Fe/H]$<$-0.4 dex,
some of them must have  already spread at various radius and therefore
contaminate  all  radii.   More  generally,  in 8  Gyr's  time,  these
processes may certainly  give rise to a complex  state of mixing among
stars  at  a  given  galactocentric  radius,  and  this  is  certainly
reflected in Fig. \ref{dfehnma}.

\subsection{Radial Migration}

Could the  detected trend be the  consequence of a  selection effect ?
The origin  of the  two first samples  are very different.   The first
sample  has not  been  selected upon  kinematic  criterion, while  the
second   is  a  compilation   of  various   spectroscopic  catalogues.
Moreover, while the two first samples are essentially contained in the
third  (CGS catalogue),  they have  only  154 stars  in common.  After
having removed these objects from the two samples, the trends shown on
Fig. 2\&3 are unaffected, which means that the correlation is observed
in two  distinct samples with different  origin, and in  a much larger
sample  (12000  stars).  We  conclude  that  it  can be  assumed  with
confidence that the trend is real.

Is there  any evidence of the  time scale of the  migration process in
the present data ? The age-metallicity distribution of Fig. 1b shows a
dearth  of  stars below  the  line  of  -0.15 dex/Gyr.  Combined  with
estimated radial  metallicity gradients (which are  uncertain and vary
roughly between 0.04  and 0.1 dex/kpc), it means  upper values for the
migration rate of  1.5 to 3.7 kpc/Gyr. Given that  the local mean disc
metallicity is [Fe/H]=0., it implies  that a star with [Fe/H]=-0.4 (or
even less) could reach the solar circle being only 2-3 Gyr old.  These
estimates are high.  Are they unrealistic ? The data are sparse in the
low metallicity  regime, and these numbers, given  as rough estimates,
need to be confirm.  However, interesting comparisons can be made with
the  few results  available from  dynamical modelling.   Binney (2007)
argued  using simple  arguments  that the  increase  of radial  random
motion by the  diffusion process described in Wielen  et al. (1996) is
not sufficient to explain  the observed scatter in the age-metallicity
distribution.  The epicycle  amplitude evaluated  by Binney  (2007) is
about 1.2 kpc,  which falls well short to explain  the presence in the
solar  neighbourhood  of stars  with  [Fe/H]=-0.6  or  even -0.8  dex.
Radial migration  based on the  effect of stochastic spiral  waves has
been  studied by  De  Simone  et al.  (2004),  giving similar  values,
limited to  at most $\pm$ 2  kpc.  However, in an  illustration of the
mechanism  described  by  Sellwood   \&  Binney  (2002),  L\'epine  et
al. (2003) concluded that stars  could have migrated radially over 2-3
kpc in  1 Gyr  or less. This  value fits  well with the  required time
scale, which of course does not  imply it is the correct mechanism. In
any  case, the  above numbers  indicate  that the  increase of  radial
random motion  does not appear sufficient to  explain radial wandering
on distances  well above 2 kpc,  and some extra-mechanism  of the kind
described by Sellwood \& Binney (2002) is required.

\begin{figure}
\includegraphics{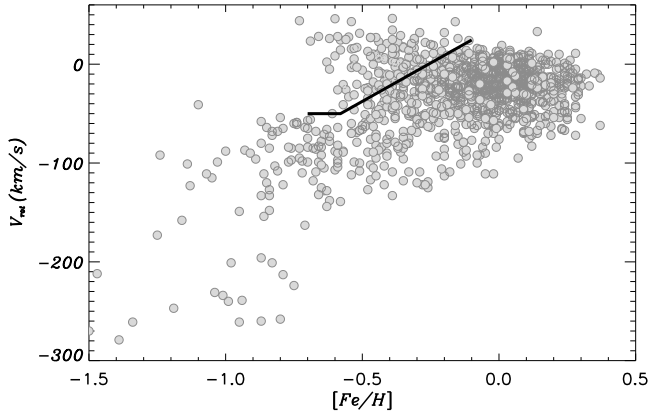}
\caption{V space velocity relative to the Sun as a function of metallicity for samples
in the solar neighbourhood from Reddy et al. (2003, 2006), Valenti \& Fischer (2005), Bensby et al. (2005)
and Gilli et al. (2006).
Metal-poor thin disc stars are clearly visible (above the line), 
as a separate group running towards low metallicity, high-V velocity component.}
\label{vrotfeh}
\end{figure}

\begin{figure}
\includegraphics{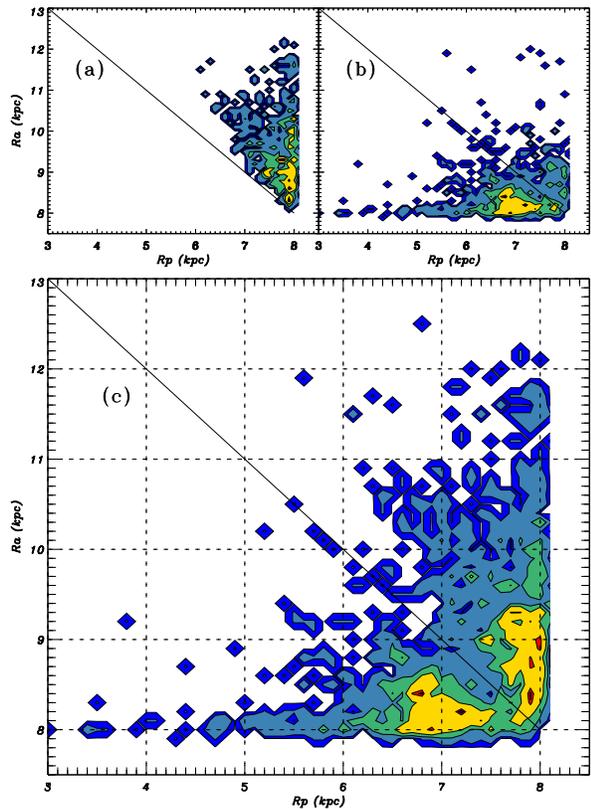}
\caption{Distribution of peri- and apo-centres (a) Stars in the GCS catalogue with 
[Fe/H]$<$-0.4 dex and V space velocity greater than -5km.s$^{-1}$. 
(b) Stars with [Fe/H]$>$+0.1 dex. 
(c) Distribution of the two subsamples (no membership probability has been used to
selected the stars on all three plots).
}
\label{radmix}
\end{figure}

\subsection{Characterising the metal-poor tail of the thin disc}

Figure \ref{vrotfeh}  shows the velocity in the  direction of galactic
rotation  relative  to  the  Sun  $vs$ metallicity  for  objects  with
detailed  spectroscopy from  Reddy  et al.  (2003,  2006), Valenti  \&
Fischer (2005), Bensby et al. (2005) and Gilli et al. (2006).  On this
figure, the  thick disc is  lagging the Sun  by about 50 to  150 km/s.
The interesting  feature is the upturn of  velocities at [Fe/H]$<$-0.4
dex for a group of stars  which rotates distinctly faster than the Sun
(to velocities up to 60  km.s$^{-1}$).  It order to characterise these
stars and illustrate their role in the systematic effects evidenced in
the previous section, we isolate  them in the survey of Nordstr\"om et
al. (2004)  by imposing [Fe/H]$<$-0.35  dex and V$>$-5  km$^{-1}$, and
plot on Figure \ref{radmix}a their distribution of apocentres $vs$ the
pericentres.  As this figure shows, the metal-poor thin disc subsample
is confined  to the  upper part  of the plot,  or mean  orbital radius
larger than 8 kpc, which is a consequence of the selection made on the
V component (V$>$-5 km$^{-1}$).  Obviously, these objects are the same
stars  that, at  low metallicities,  skew the  mean orbital  radius of
Fig. 2 and 3 above the mean.

The  comparison   with  the   distribution  of  metal-rich   stars  on
Fig.  \ref{radmix}(b) is  suggestive :  the two  groups  occupy almost
separate,    complementary   areas    in   the    (R$_{p}$,   R$_{a}$)
plane. Fig.  \ref{vrotfeh}(c) shows both groups,  with the symmetrical
distribution  around   the  mean  orbital   radii  at  8   kpc  neatly
illustrated.    The   group   of   metal-poor  stars   identified   in
Fig.  \ref{dfehnma}, \ref{vrotfeh}  and  \ref{radmix} are  undoubtedly
thin disc objects.  This is supported by the  UVW velocity dispersions
measured on the 572 stars  of our selection: (32, 11, 21) km.s$^{-1}$,
and it is confirmed by the ages and $\alpha$-element content discussed
in section  3.  These stars are  the signatures that  radial mixing in
the disc is an efficient process.

\begin{figure}
\includegraphics{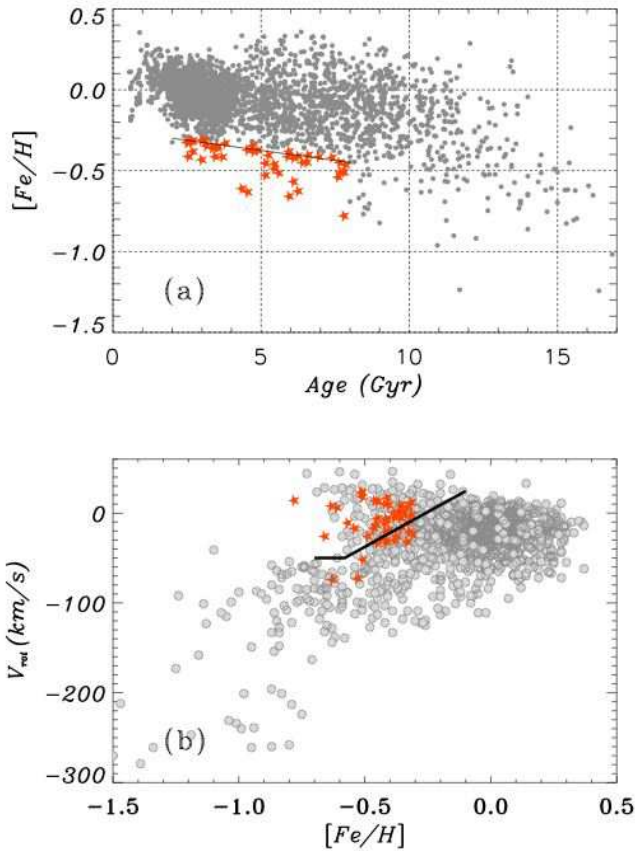}
\caption{Stars responsible for the large dispersion towards low metallicities are selected 
from the age-metallicity distribution in panel (a) in the age range 2-8 Gyr. 
(b) Their corresponding location in the V-[Fe/H] plot.}
\label{agemetwand}
\end{figure}

As  expected,  they are  also  responsible  for  the large  dispersion
towards  low metallicity  at a  given age.  This is  testified  by the
Fig. \ref{agemetwand}ab, where we selected stars of the thin disc with
low  metallicities   from  the  photometric   sample  (age-metallicity
distribution  of panel  a).  Their  position in  the  V-[Fe/H] diagram
corresponds to the  same location occupied by the  group of metal-poor
thin disc stars identified on Fig. \ref{vrotfeh} and \ref{radmix}.

\section{The transition between the thick and thin discs}

The results above have several consequences on the way we understand the metallicity
distribution of thin disc stars, and the transition between the thin and the thick 
discs, as we now discuss.

\subsection{Solving the problem of the metallicity hiatus between the two discs}

\begin{figure*}
\includegraphics{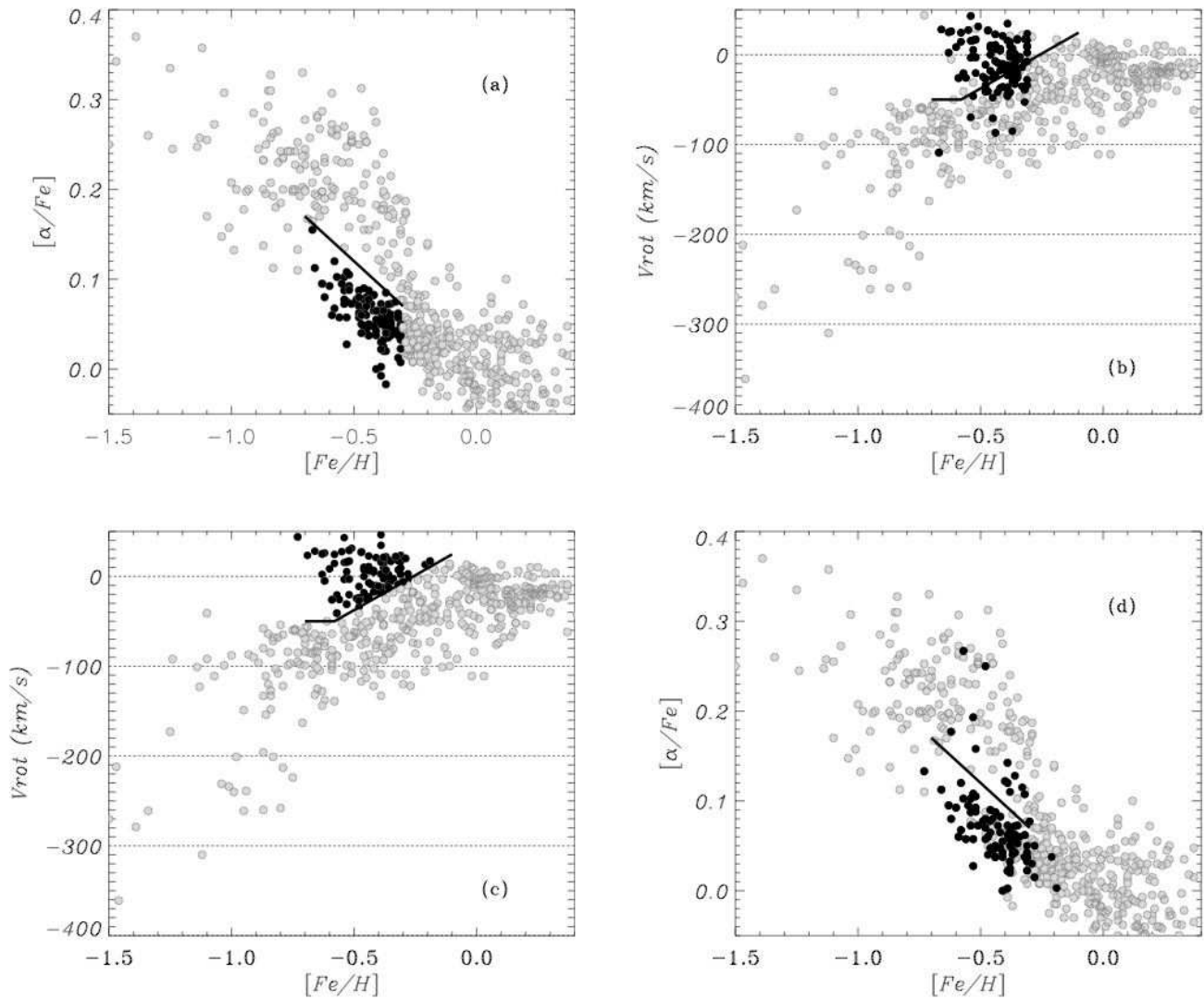}
\caption{(a) [$\alpha$/Fe] $vs$ [Fe/H] for stars in the samples of 
 Reddy et al. (2003,2006), Bensby et al. (2005) and Gilli et al. (2006).
Black dots highlight the selection made for stars with [$\alpha$/Fe] below the line
and -0.7$<$[Fe/H]$<$-0.3 dex. (b). V space velocity (wrt the Sun) from Nordstr\"om et al. (2004)
as a function of metallicity for the same samples, and location of objects selected in panel (a).
Panel (c) : V velocity $vs$ metallicity. Objects are now selected (above the line) 
on the basis of metallicity and V velocity. 
Panel (d) : location of selected objects (panel c) on the [$\alpha$/Fe] $vs$ [Fe/H] diagram.
To the exception of a few outliers, the group of stars responsible for the hiatus in
metallicity is also the group which rotate faster than the Sun.
}
\label{alphafeh}
\end{figure*}

We first focus on the hiatus in metallicity between the two discs. 
How does radial mixing of stars in the disc can help to solve this problem ?
In Fig. \ref{alphafeh}(a) we display [$\alpha$/Fe] and [Fe/H] values 
from the samples of Reddy et al. (2003, 2006), Bensby et al. (2005) and Gilli et al. (2006).
Black dots represent stars that have a metallicity between -0.7 and -0.3 dex, 
and [$\alpha$/Fe] below the line, selecting metal-poor
thin disc stars. The location of these objects on the V$_{rot}$ $vs$ [Fe/H] diagram 
in Fig. \ref{alphafeh}(b)
illustrates that the metal-poor stars responsible for the hiatus in metallicity 
are the same group of stars studied in the previous section and which rotate faster than the Sun. 
In other words, if 
our tentative interpretation of section 2.3 is correct, they are wanderers from the outer disc.
About 30\% of the stars selected in panel (a) fall below the line in panel (b), which means
that for a few cases, this line is too simplistic to differentiate accurately the kinematics of metal-poor
thin stars from the main population. In some cases, the velocity in the direction of 
galactic rotation is clearly in the thick disc regime. 
This is not unexpected, since dynamical processes may have mixed 
the two groups to some extent. 

Panel (c) and (d) below illustrate the location of the stars in the [$\alpha$/Fe] $vs$ [Fe/H] 
diagram when the selection is made on the V velocity component 
and metallicity. We selected stars of the metal-poor
thin disc in panel (c) above the line. Panel (d) 
shows that this simple selection corresponds essentially to stars in the group 
responsible for the hiatus in metallicity. About 15\%  
of the stars selected this way are 
not in the area of metal-poor thin disc stars of panel (d) (and appear as transition or
thick disc stars). Here again, this is not unexpected, since the selection is made simple and kinematic
evolution certainly have mixed the two populations to some extent.

Fig. \ref{alphafeh} demonstrate that
the objects responsible for the hiatus in metallicity seen between the two populations probably
originate from the outer disc. These objects have kinematics, 
metallicity and, as we show in section 3.3, age distribution incompatible with being transition objects 
between the thick and thin discs, which imply that the hiatus in metallicity is only apparent, and
is not an effect of the local chemical evolution.

\subsection{Transition stars}

\begin{figure}
\includegraphics[width=8.5cm]{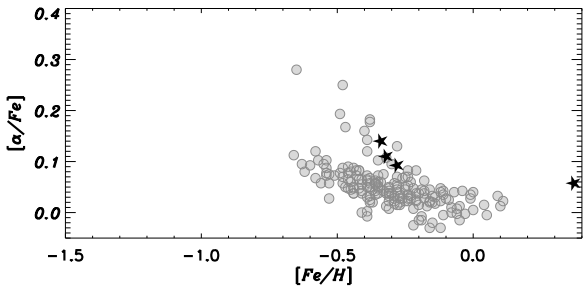}
\caption{Distribution of stars kinematically selected as thin disc stars in Reddy et al. (2006, their Fig. 12).
Two branches are clearly visible, one made of metal-poor thin disc stars, the other of transition objects.
Additional transition objects (as defined by kinematic criteria only) from Bensby et al. (2005) are also plotted (star symbols).
}
\label{tranobject}
\end{figure}

We may expect transition stars, if a transition phase between the two
discs does  exist, to reach  the thin disc  at a metallicity  of about
[-0.2,-0.3] dex,  spread at  various  level  of $\alpha$-element  enrichment
between the two populations.   Inspection of the literature shows that
these objects  are relatively rare.   Partly, this is  because adopted
kinematic criteria in current studies are sometimes chosen to select
stars with high probability to  belong to one or the other population,
in order to enhance the differences in the chemical properties of the
two populations.  This may bias samples  against transition objects,
if  their kinematics is not clearly in the thin disc or thick disc group. 
The  main  reason however  is
probably that they are intrinsically rare.  Bensby et al. (2005) found
4  (kinematically  defined) transition or 'intermediate' objects 
(meaning with no clear thin or thick disc kinematics).   Their  stars  have
$\alpha$-elements well  intermediate between the thin  and thick discs
(see their  Fig. 8).  The suggestive clue  here is  their metallicity,
which for  3 of them are  within [-0.34, -0.28]  (one has [Fe/H]=+0.37
dex), well above  the lower bound metallicity of  metal-poor thin disc
stars  (at -0.6,-0.8  dex) discussed  here.  This  is well  within our
estimated value for  the transition between the two  discs.  A similar
indication is given by the transition stars of Fuhrmann (2004, see his
Fig. 34 and 35).  Finally, the  clearest example is seen on Fig. 12 of
Reddy  et al.  (2006),  which  shows [$\alpha$/Fe]  as  a function  of
[Fe/H], reproduced here in Fig. \ref{tranobject}.  The data of Reddy et
al. (2003) and (2006) suggest  that, when transition or disc stars are
(kinematically) selected (panel b and  c on their figure), the pattern
of $\alpha$-elements  shows two branches.  The first one  goes towards
the metal-poor stars of the  thin disc discussed previously. The other
is starting  at [Fe/H]=(-0.2,-0.3) dex and  [$\alpha$/Fe]=0.05 dex and
aims  at the  thick  disc ([$\alpha$/Fe]=0.2  dex  and metallicity  of
[Fe/H]=(-0.4,-0.5) dex).   These are robust indices  suggesting that a
transition  phase  between the  two  populations  has  occurred in  a
relatively  narrow  range  of  metallicity.   It is  possible  that  a
particularly low  level of star formation  activity characterises this
transition  phase, although  this is  difficult assess,  in particular
because the duration of this phase is difficult to estimate.

\subsection{Ages: transition and metal-poor thin disc stars}

\begin{figure}
\includegraphics{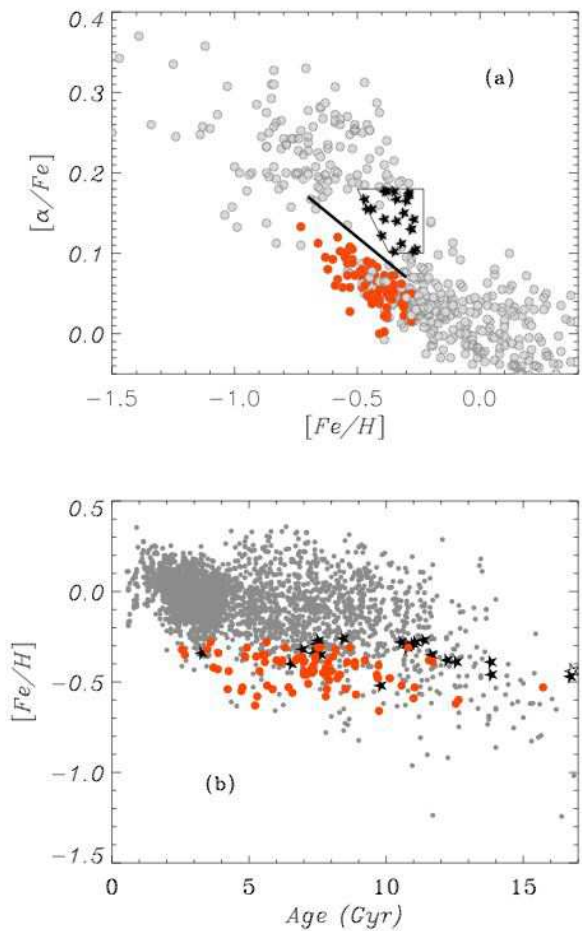}
\caption{(a) [$\alpha$/Fe] as a function of metallicity for stars in the samples of 
Reddy et al. (2003, 2006) 
and Bensby et al. (2005). Two subsamples have been selected as detailed in the text. The first, 
represented by (orange) dots, are metal-poor thin disc objects. The second, represented by star symbols
are transition stars. 
(b) The position of these
two groups in the age-metallicity distribution. The open star symbol is for a star (HIP 65449) which age 
was found greater than 17 Gyr.
}
\label{alphafehvrot}
\end{figure}

We now comment on the age of the stars in both groups (metal-poor thin
disc  and  transition stars).   The  metal-poor  thin  disc stars  are
expected to  have any  age between a  few Gyr  (the time they  need to
approach the solar  circle) and the age of  the disc, while transition
stars are expected to be exclusively  old, that is, older than the age
of  the thin  disc at  the solar  circle.  Individual  ages  have been
determined for  the samples  of Reddy et  al. (2003, 2006),  Bensby et
al.  (2005) and  Gilli et  al. (2006)  using the  method  described by
J{\o}rgensen  \& Lindegren  (2005), for  stars with  M$_{v}<$5.0.  
Metallicities,  $\alpha$-element content and effective temperatures of the stars  are taken from
the studies  above. Absolute magnitudes are taken from the Hipparcos catalogue.
We selected metal-poor  stars of the thin disc imposing both
criteria defined in Fig.  \ref{alphafeh}ac, while transition stars were
chosen visually  from the box  in Fig. \ref{alphafehvrot}a.   There is
actually  no working  definition of  what  a transition  star is.   In
principle, transition stars  ought to be born at  an intermediate epoch
between the  thick and thin discs,  but we don't  know what measurable
property would select stars as  near as possible from this definition.
Kinematically   intermediate   stars don't   fulfill   this
requirement with  high precision. As  an example, stars  with asymmetric drift
intermediate between the thin and thick discs in  Reddy et al. (2006) have  metallicity running from
-0.87 to +0.37  and [$\alpha$/Fe] from -0.05 to  0.24 dex, which means
they  are not  specifically intermediate  in terms  of  chemistry.  In
spite  of the  relatively small  dependence of  age on  metallicity, a
combination of  the $\alpha$-element content and  metallicity seems to
be a much better proxi than kinematics.  We therefore select, somewhat
arbitrarily, transition  stars as objects  that are within  the region
delimited on Figure \ref{alphafehvrot}a.  Panel (b) shows the position
of both samples (metal-poor thin disc and transition stars) superposed
to the age-[Fe/H]  distribution of Fig. \ref{rmeffect}b.  The age-scale
is the same for all three samples (determined by the same method using the
same set of isochrones).

The respective locations of thin disc metal-poor and transition stars confirm that transition 
stars are, in general, older than thin disc objects. 
The dichotomy between the two  groups is apparent, and corresponds to
our expectations. Metal-poor thin disc objects are spread at all ages
starting from a few Gyr ($>$ 2 Gyr) to about 10 Gyr for most stars.
67\% of thin disc metal-poor stars have ages less than 8 Gyr, while another 21 \% have error bars
which make them compatible with this limit.
Transition stars are mostly older than 6 Gyr. The group of transition stars at ages near 7-8 Gyr
have error bars making them compatible  with being older than 8 Gyr. 
There is one outlier, HIP 26828, which has
([Fe/H],[$\alpha$/Fe])=(-0.34, 0.16) dex and
seems to be truly intermediate in terms of chemistry and kinematics. It is positioned 
in the overlap region in the HR diagram, and its age pdf indicates two peaks, one 
at 3.3 Gyr, the other at 4.5 Gyr.  In any case, its age looks incompatible with 
the other properties, and it appears as a peculiar object 
in this distribution (a blue straggler ?). 

\begin{figure}
\includegraphics[]{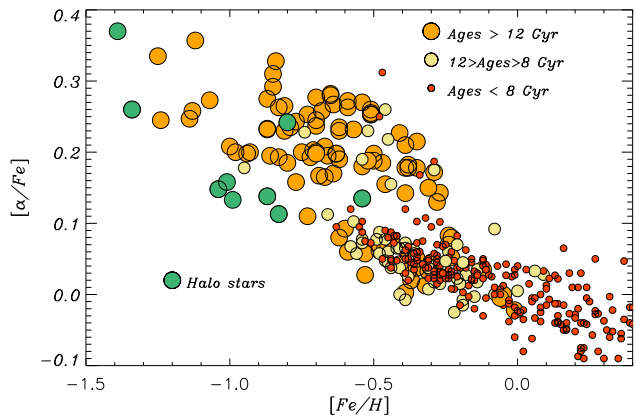}
\caption{Age distribution of stars in the  [$\alpha$/Fe] $vs$ [Fe/H] diagram from Reddy et al. (2003, 2006),
Bensby et al. (2005) and Gilli et al. (2006). 
Note the dichotomy between thick disc stars ([$\alpha$/Fe]$>$+0.18), which mostly have 
ages $>$ 12 Gyr and the solar metallicity ($\pm$0.2 dex) thin disc, with most ages under 8 Gyr.
}
\label{alphafehage}
\end{figure}

\begin{figure}
\includegraphics[]{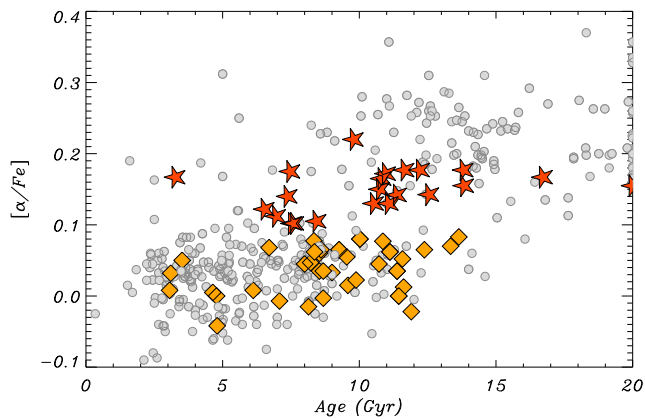}
\caption{[$\alpha$/Fe] as a function of age for the same stars. Transition objects of Fig. 10 
are shown as star symbols. Metal-poor thin disc objects of Fig. 10 have not been
plotted. Diamonds represent objects that have kinematics in the thick disc regime
and [$\alpha$/Fe]$<$0.1 dex. The age of most of these objects is 
intermediate between the two discs, at 8-12 Gyr.
}
\label{alphage}
\end{figure}

The  age information  is summarised  in Fig.  \ref{alphafehage}, where
stars are represented in three different age intervals.  If we discard
thin  disc  metal-poor  stars  as   not  being  stars  born  at  solar
galactocentric     distance,    the     age-ranges     displayed    on
Fig. \ref{alphafehage} shows that the thick disc is dominated by stars
being older than  12 Gyr, while the thin disc  seems to be essentially
younger than 8  Gyr at [Fe/H]$>$-0.25 dex.  A  different view is given
on Fig. \ref{alphage}, which shows the same data with [$\alpha$/Fe] as
a function of  age. The metal-poor thin disc stars  (shown on Fig. 10)
have been  removed from  the plot,  in order to  keep only  stars that
would be  truly endemic of  the solar neighbourhood, according  to the
scheme presented  in previous sections.   Transition stars of  Fig. 10
are shown as star symbols. In addition, we show as diamonds stars that
have been selected with the conditions that [$\alpha$/Fe]$<$0.1 dex, V
space velocity component less than -40 km.s$^{-1}$, and total velocity
greater than  80 km.s$^{-1}$.   For all categories,  there is  a large
spread in ages,  but the mean increase of  [$\alpha$/Fe] as a function
of age  is apparent.  Objects  shown as diamonds are  kinematically in
the thick  disc regime, while  their [$\alpha$/Fe] abundance  ratio is
more akin to the thin disc.  According to their age, these objects are
mostly old  thin disc,  with a majority  having 8-12 Gyr.  While their
status is difficult to define, a possible interpretation is that these
objects should perhaps be considered as transition objects between the
two  discs, but on  the side  of thin  disc, while  transition objects
discussed  in section  3 are  transition objects  nearer to  the thick
disc.

\section{Summary and discussion}

\subsection{The thin disc}

Analysis of the  orbital parameters of local samples  of stars shows a
systematic correlation of the mean orbital radius of thin disc stars 
with  metallicity. It  is  shown that  stars in  the
metal-poor ([Fe/H]$<$-0.3 dex) and metal-rich ([Fe/H]$>$+0.2 dex) tails
of  the  thin disc  have  orbital  parameters  significantly off  the  main
population, suggesting an origin in the outer and inner galactic disc.

We note that local low-metallicity  ([Fe/H]$<$-0.3 dex)  thin-disc  stars have
properties  well  in agreement  with  stars  found  in the  anticentre
direction  a  few  kpc   from  the  solar  neighbourhood.   Carney  et
al. (2005)  and Yong et al.  (2005) find that the  metallicity of thin
disc stars decreases  towards the outer disc to  reach a constant value
of  [Fe/H]$\approx$-0.6  dex  2-4 kpc beyond the solar circle.   Moreover,  the  ratio
[$\alpha$/Fe]$\approx$  0.1 to  0.2  dex, as  evaluated  by Carney  et
al. (2005)  on their stars is compatible  with similar characteristics
of the local thin disc stars at [Fe/H]$<$-0.5 dex (i.e slightly above thin
disc stars of solar metallicity).   These values are in agreement with
the metal-poor  thin disc sampled  locally.  

An important  characteristic of metal-poor thin disc  objects is their
distinctly  higher-than-average  space velocity  in  the direction  of
galactic  rotation.   Their location  in  the ([$\alpha$/Fe],  [Fe/H])
diagram shows that they are  responsible for the hiatus in metallicity
between  the  two  discs.   Three other  important  consequences  that
concern the local chemical evolution must be noted.  First, when their
distinct provenance is taken into account, stars of the thin disc that
are  truly  endemic  of the  solar  circle  span  a limited  range  in
metallicity (from  about -0.2 to 0.2  dex).  This is  confirmed by the
mean evolution of the metallicity  within the thin disc, which remains
remarkably   flat  over   8-10   Gyr.   Second,   when  sampling   the
age-metallicity distribution of the solar vicinity, care must be taken
to select stars  that were genuinely formed at  the solar circle.  For
instance, the  stars in the  sample of Reddy  et al. (2003)  have been
selected (voluntarily in  this case) by the authors  to best represent
metal-poor   thin  disc.   It   implies  that   their  age-metallicity
distribution (their Fig. 8) is  representative of the outer thin disc,
and is offset, at a given age, to lower metallicities when compared to
stars truly  endemic of the solar galactocentric  radius.  Finally, in
discussing  the G-dwarf  problem, the  amount of  the  missing stellar
material at low metallicity (compared to the closed-box model) resides
entirely  in the thick  disc metallicity  regime.  The  missing dwarfs
have typical metallicity of 1/3 of  solar, or about -0.5 dex.  This is
clearly outside the range of the local thin disc.

\subsection{The thick  disc status}

If the  above analysis  is correct, a  real evolutionary link  is seen
between the two disc components, and  it is difficult to imagine how a
stellar thick  disc of purely  extragalactic origin could  insure such
parenthood.   It must  be emphasised  however that  the status  of the
thick disc, and  of the stars labeled as  such in recent spectroscopic
studies  of local  samples, is  becoming increasingly  ambiguous.  The
recent  literature   conveys  the  impression   that  this  population
possesses complex or even  contradicting properties, either locally in
the  solar neighbourhood  or  on  a global  footage,  when looking  at
in-situ data.   On the one side,  studies of in-situ samples  have claimed
the detection  of structures  (Gilmore, Wyse \&  Norris 2002,  Wyse et
al. 2006) although their  characterisation and identification with the
thick   disc  relics   remains   challenging.  On   the  other   side,
star-counting  studies seem  to acknowledge  a 'regular'  thick disc,
although  it must  be recognised  that  these studies  have failed  to
characterise  uniquely  the properties  of  this  population on  large
scales (see  for example Chen  et al. 2001),  and one may ask  to what
extent possible structures are responsible for this.

It is neither clear from local samples what exactly the thick disc is.
Helmi  et  al.  (2006),  Arifyanto  \&  Fuchs  (2006)  and  Bensby  et
al. (2007)  all find conclusive evidence for  kinematical groupings in
local samples  in the  range of metallicity  and kinematics  where the
thick  disc is  likely  to dominate.   These  studies illustrate  that
stellar groups are  found lagging the local standard of rest at  various 
speed from about
-125 km.s$^{-1}$  (Arcturus stream, Navarro, Helmi  \& Freeman, 2004),
-80km.s$^{-1}$ (Arifyanto \& Fuchs, 2006) to -50 km.s$^{-1}$ (Hercules
stream, Bensby  et al.,  2007, Soubiran \&  Girard, 2005), and  one is
justified  to ask:  what room  remains for  a 'regular'  thick  disc ?
Since stellar streams are found lagging every 30 to 40 km/s behind the
old  thin  disc,   how  do  we  know  that   stars  sampled  by  local
spectroscopic surveys are tracing  a supposedly existing regular thick
disc  ?   Or do  we  have  to assume  that  the  local  thick disc  is
essentially an accumulation of stellar streams ?

Additional confusion  comes from the  fact that stars  labelled 'thick
disc'  by kinematic  membership probability  in  local spectroscopic
surveys are far from the canonical characteristics of this population.
Thick disc stars selected  upon [$\alpha$/Fe]$>$0.2 dex in all samples
studied here have a mean lag with respect to the Sun of about -74 km.s$^{-1}$
(-85  in Soubiran  \&  Girard 2005).   This  is to  contrast with  the
canonical   value  of   the   asymmetric  drift   assumed  to   derive
probabilities,  which are -46  km.s$^{-1}$ (Bensby  et al.  2005), -48
km.s$^{-1}$ (Reddy et al.  (2003, 2005), and -51 km.s$^{-1}$ (Soubiran
\& Girard, 2005).  Similarly, the probability membership, when applied
to the GCS catalogue, yields  324 stars with probabilities higher than
80\%  to belong  to this  population, with  a mean  rotational  lag of
-83km.s,    and   mean    metallicity   of    -0.46   dex.     
This value of the rotational lag of the thick disc do bear some resemblance
with   the  component   found  by   Arifyanto  \&   Fuchs   (2006)  at
V$_{lag/sun}$=-85km/s, with  a metallicity distribution  that seems to
be  highly  dispersed  and   asymmetrical,  with  most  of  the  stars
concentrating in the range [-1.0,  0.0] dex.  Similarly, at least part
of  the  structures identified  by  Helmi  et  al. (2006)  are  likely
contained  in  the 'thick  disc'  we  sampled  from the  catalogue  of
Nordstr\"om  et al. (2004).   In any  case, these  data sets  reveal a
population which  is rather  far from the  canonical thick  disc.  The
general  consequence   of  these  studies  has   been  the  increasing
complexity  in the descriptions  of the  local data,  without actually
achieving  a clear  identification  of the  thick  disc.  The  overall
picture  seems  to  point   toward  a  lumpy  component,  with  direct
connections with the thin disc,  but the picture is fragmentary, and a
consistent description of the  Galaxy at intermediate metallicities is
still lacking.

\section*{Acknowledgments}

 
\bsp


\begin{thebibliography}{}

\bibitem[\protect\citeauthoryear{Arifyanto \& 
Fuchs}{2006}]{2006A&A...449..533A} Arifyanto M.~I., Fuchs B., 2006, A\&A, 
449, 533 

\bibitem[\protect\citeauthoryear{{Bensby}, {Feltzing}, \& {Lundstr{\"
  o}m}}{{Bensby} et~al.}{2003}]{03BEN_EA}
{Bensby} T., {Feltzing} S.,  {Lundstr{\" o}m} I., 2003, A\&A, 410, 527

\bibitem[\protect\citeauthoryear{Bensby et al.}{2005}]{2005A&A...433..185B} 
Bensby T., Feltzing S., Lundstr{\"o}m I., Ilyin I., 2005, A\&A, 433, 185 

\bibitem[\protect\citeauthoryear{Bensby et al.}{2007}]{2007ApJ...655L..89B} 
Bensby T., Oey M.~S., Feltzing S., Gustafsson B., 2007, ApJ, 655, L89 

\bibitem[\protect\citeauthoryear{Binney}{2007}]{2007iuse.book...67B} Binney 
J., 2007, Island Universes,p 67, Springer, Astrophysics and Space Science Proceedings

\bibitem[\protect\citeauthoryear{Carney et al.}{2005}]{2005AJ....130.1111C} 
Carney B.~W., Yong D., de Almeida M.~L.~T., Seitzer P., 2005, AJ, 130, 1111 

\bibitem[\protect\citeauthoryear{Chen et al.}{2001}]{2001ApJ...553..184C} 
Chen B., et al., 2001, ApJ, 553, 184 

\bibitem[\protect\citeauthoryear{Demarque et 
al.}{2004}]{2004ApJS..155..667D} Demarque P., Woo J.-H., Kim Y.-C., Yi 
S.~K., 2004, ApJS, 155, 667 

\bibitem[\protect\citeauthoryear{De Simone, Wu, \& 
Tremaine}{2004}]{2004MNRAS.350..627D} De Simone R., Wu X., Tremaine S., 
2004, MNRAS, 350, 627 

\bibitem[\protect\citeauthoryear{Edvardsson et 
al.}{1993}]{1993A&A...275..101E} Edvardsson B., Andersen J., Gustafsson B., 
Lambert D.~L., Nissen P.~E., Tomkin J., 1993, A\&A, 275, 101 

\bibitem[\protect\citeauthoryear{Famaey et al.}{2005}]{2005A&A...430..165F} 
Famaey B., Jorissen A., Luri X., Mayor M., Udry S., Dejonghe H., Turon C., 
2005, A\&A, 430, 165

\bibitem[\protect\citeauthoryear{Famaey et al.}{2007}]{2007A&A...461..957F} 
Famaey B., Pont F., Luri X., Udry S., Mayor M., Jorissen A., 2007, A\&A, 
461, 957 

\bibitem[\protect\citeauthoryear{Feltzing, Holmberg, \& 
Hurley}{2001}]{2001A&A...377..911F} Feltzing S., Holmberg J., Hurley J.~R., 
2001, A\&A, 377, 911 

\bibitem[\protect\citeauthoryear{Fuhrmann}{1998}]{1998A&A...338..161F} 
Fuhrmann K., 1998, A\&A, 338, 161 

\bibitem[\protect\citeauthoryear{Fuhrmann}{2004}]{2004AN....325....3F} 
Fuhrmann K., 2004, AN, 325, 3

\bibitem[\protect\citeauthoryear{Gilmore, Wyse, \& 
Norris}{2002}]{2002ApJ...574L..39G} Gilmore G., Wyse R.~F.~G., Norris 
J.~E., 2002, ApJ, 574, L39 

\bibitem[\protect\citeauthoryear{Grenon}{1972}]{1972ade..coll...55G} Grenon 
M., 1972, in  G. Cayrel de Strobel and A. M. Delplace, eds, 
Age des Etoiles, Proceedings of IAU Colloq. 17, Paris, France, p 55.

\bibitem[\protect\citeauthoryear{{Haywood}}{{Haywood}}{2002}]{02HAY}
{Haywood} M., 2002, MNRAS, 337, 151

\bibitem[\protect\citeauthoryear{{Haywood}}{{Haywood}}{2006}]{02HAY}
{Haywood} M., 2006, MNRAS, 337, 151


\bibitem[\protect\citeauthoryear{Helmi et al.}{2006}]{2006MNRAS.365.1309H} 
Helmi A., Navarro J.~F., Nordstr{\"o}m B., Holmberg J., Abadi M.~G., 
Steinmetz M., 2006, MNRAS, 365, 1309 


\bibitem[\protect\citeauthoryear{J{\o}rgensen \& 
Lindegren}{2005}]{2005A&A...436..127J} J{\o}rgensen B.~R., Lindegren L., 
2005, A\&A, 436, 127 

\bibitem[\protect\citeauthoryear{L{\'e}pine, Acharova, 
\& Mishurov}{2003}]{2003ApJ...589..210L} L{\'e}pine J.~R.~D., Acharova I.~A., Mishurov Y.~N., 2003, ApJ, 589, 210 

\bibitem[\protect\citeauthoryear{Minchev \& 
Quillen}{2006}]{2006MNRAS.368..623M} Minchev I., Quillen A.~C., 2006, 
MNRAS, 368, 623 

\bibitem[\protect\citeauthoryear{Mishenina et 
al.}{2004}]{2004A&A...418..551M} Mishenina T.~V., Soubiran C., Kovtyukh 
V.~V., Korotin S.~A., 2004, A\&A, 418, 551 

\bibitem[\protect\citeauthoryear{Navarro, Helmi, \& 
Freeman}{2004}]{2004ApJ...601L..43N} Navarro J.~F., Helmi A., Freeman 
K.~C., 2004, ApJ, 601, L43 

\bibitem[\protect\citeauthoryear{{Nordstr{\" o}m} et~al.}{{Nordstr{\" o}m}
  et~al.}{2004}]{04NOR_EA}
{Nordstr{\" o}m} B., Mayor, M., Andersen, J., et~al., 2004, A\&A, 418, 989

\bibitem[\protect\citeauthoryear{Reddy et al.}{2003}]{2003MNRAS.340..304R} 
Reddy B.~E., Tomkin J., Lambert D.~L., Allende Prieto C., 2003, MNRAS, 340, 
304 

\bibitem[\protect\citeauthoryear{Reddy, Lambert, \& Allende 
Prieto}{2006}]{2006MNRAS.367.1329R} Reddy B.~E., Lambert D.~L., Allende 
Prieto C., 2006, MNRAS, 367, 1329 

\bibitem[\protect\citeauthoryear{Sellwood \& 
Binney}{2002}]{2002MNRAS.336..785S} Sellwood J.~A., Binney J.~J., 2002, 
MNRAS, 336, 785

\bibitem[\protect\citeauthoryear{Soubiran \& Girard}{2005}]{2005A&A...438..139S} Soubiran C., Girard P., 2005, A\&A, 438, 139 

\bibitem[\protect\citeauthoryear{Wielen, Fuchs, \& 
Dettbarn}{1996}]{1996A&A...314..438W} Wielen R., Fuchs B., Dettbarn C., 
1996, A\&A, 314, 438 

\bibitem[\protect\citeauthoryear{Wielen}{1977}]{1977A&A....60..263W} Wielen 
R., 1977, A\&A, 60, 263 

\bibitem[\protect\citeauthoryear{Wyse et al.}{2006}]{2006ApJ...639L..13W} 
Wyse R.~F.~G., Gilmore G., Norris J.~E., Wilkinson M.~I., Kleyna J.~T., 
Koch A., Evans N.~W., Grebel E.~K., 2006, ApJ, 639, L13 

\bibitem[\protect\citeauthoryear{Yong, Carney, \& de 
Almeida}{2005}]{2005AJ....130..597Y} Yong D., Carney B.~W., de Almeida 
M.~L.~T., 2005, AJ, 130, 597 

\end{thebibliography}
\end{document}